# Longitudinal Seebeck coefficient of the charge-ordered layered crystals in a strong quantizing magnetic field

P.V. Gorsky

Institute of Thermoelectricity of the National Academy of Sciences and Ministry of Education and Science, Youth and Sports of Ukraine, 58001, Chernivtsi, Ukraine

**Summary**

The longitudinal Seebeck coefficient of the charge-ordered layered crystals in a strong quantizing magnetic field normal to layers plane has been determined.

The conditions whereby charge ordering parameter and chemical potential of charge carriers are the oscillating functions of the magnetic field induction are considered. The longitudinal Seebeck coefficient has been calculated for two models of the relaxation time: i) constant relaxation time and ii) the relaxation time proportional to the longitudinal velocity. It has been shown that in a quasi-classical region of magnetic fields for the case of the relaxation time proportional to the longitudinal velocity the longitudinal Seebeck coefficient is less than for the case of constant relaxation time. In this region, for selected problem parameters it does not exceed 4.37μV/K. In the strong quantizing magnetic fields for both models of the relaxation time the longitudinal Seebeck coefficient is virtually the same. For selected problem parameters its maximal modulus is 2033μV/K. At the same time, in the disordered layered crystals, in a quasi-classical region, the Seebeck coefficient is approximately one order of magnitude less than for the charge ordered crystals. In the strong magnetic fields, the Seebeck coefficient for the disordered layered crystals is factor of 7 to 9 less than for the charge-ordered crystals. However, in super strong magnetic fields, under current carriers concentration in the only filled Landau sub-band, for both models of the relaxation time the modulus of the Seebeck coefficient tends to zero according to asymptotic law $\alpha_{zz} \propto B^{-2}$.

1. **Introduction**

At the present time much attention is given to the development and study of the properties of new thermoelectric materials. The objects of experimental investigation are metals, alloys, semiconductors [1, 2], fullerenes [3], composites [4], including biomorphic [5], etc. Theory of thermoelectric properties of materials, including nanosystems, is being actively developed as well [6, 7]. One of the first works on the theory of transverse Seebeck coefficients of metals in quantizing magnetic fields was performed by Kosevich and Andreyev [8].

Many of the investigated materials, for instance, semiconductor systems of $A^{II}B^{VI}C^{VII}$ class, intercalated graphite compounds, synthetic metals, graphene, etc. belong in their crystal structure to layered materials. At the same time, the overwhelming majority of theoretical works dedicated to behaviour of such layered systems in quantizing magnetic fields are mainly concerned with the transverse galvanomagnetic effects. The author of this paper is aware of only one work which deals with the thermal conductivity of graphene in a quantizing magnetic field [9]. In so doing, its Fermi surface is considered to be open, that is, such which occupies the entire one-dimensional Brillouin zone and, with a periodic continuation, is a connected one, that is, represents a continuous corrugated cylinder.

At the same time, in many layered compounds, charge ordering of the type of charge-density waves (CDW) was observed whose wave vector can both lie in the layers plane and be normal to it. In one of his previous works [10] the author of this paper showed that interlayer charge ordering is typical of such layered crystals whose conductivity anisotropy under structural phase transition of CDW type increases and, in so doing, charge ordering produces no or weak effect on the low-temperature electron heat capacity of crystal away from the transition point. Transition into the charge-ordered state in such crystals can be manifested as a topological one, i.e. such whereby the Fermi surface changes connectivity, being converted from a closed surface into an open one.

In this connection, the purpose of this paper is to study the effect of charge ordering and topological transition from an open surface to a closed one due to a quantizing magnetic field on the longitudinal Seebeck coefficient of a layered crystal in the case when the quantizing magnetic field and the temperature gradient are parallel to each other and normal to the layers.

**2. Research on the longitudinal Seebeck coefficient of the charge-ordered layered crystals in a quantizing magnetic field**

The energy spectrum of conduction electrons of a layered crystal in a quantizing magnetic field normal to layers can be represented as follows:

$$\varepsilon(n,x) = \mu^* B(2n+1) + W(x), \qquad (1)$$

where $\mu^* = \mu_B m_0 / m^*$, $\mu_B$ — the Bohr magneton, $m^*$ — effective electron mass in the layers plane, $B$ — magnetic field induction, $n$ — the Landau level number, $W(x)$ — the law of electrons dispersion in the direction normal to layers, $x = ak_z$, $k_z$ — a quasi-pulse component in the direction normal to layers, $a$ — the distance between translation equivalent layers.

The use of kinetic Boltzmann equation yields the following general formula for the longitudinal Seebeck coefficient:

$$\alpha_{zz} = \frac{\sum_\beta \tau_\beta v_{z\beta}^2 \dfrac{\partial f^0(\varepsilon_\beta)}{\partial T}}{e \sum_\beta \tau_\beta v_{z\beta}^2 \dfrac{\partial f^0(\varepsilon_\beta)}{\partial \zeta}}. \qquad (2)$$

In this formula, $e$ — electron charge, $\beta \equiv (n,x)$, $\tau_\beta$ — relaxation time, $v_{z\beta}^2$ — square of electron longitudinal velocity, $f^0(\varepsilon_\beta)$ — the Fermi-Dirac distribution function, $\varepsilon_\beta$ — electron energy, $\zeta$ — chemical potential of electron system, $T$ — absolute temperature. Summation over the Landau levels in this formula can followed out in the case when relaxation time depends only on the longitudinal quasi-pulse. It is done in the same way as in [11].

This paper, just as [11], deals with the simplest model of interlayer charge ordering within the framework of which it is a simple alternation of layers with different electron density in conformity with the formula:

$$n_i = n_0 a \left[1 + (-1)^i \delta\right], \qquad (3)$$

In this formula, $i$ – layer number, $n_0$ – average volumetric concentration of charge carriers, $a$ – superlattice parameter, $\delta$ – order parameter describing the nonuniformity of filling the layers with electrons, when $0 \leq \delta \leq 1$. The case $\delta = 1$ corresponds to full ordering, and the case $\delta = 0$ – to fully disordered state.

The law of dispersion $W(x)$ for the charge-ordered layered crystal is of the form [11]:

$$W(x) = \pm\sqrt{W_0^2 \delta^2 + \Delta^2 \cos^2 x}. \qquad (4)$$

In this formula, $W_0$ – effective attracting interaction caused by competence of electron-phonon interaction with phonon mode whose wave vector is equal to the wave vector of charge density wave (CDW), i.e. $\pi/a$ and the Coulomb repulsion of electrons leading to interlayer charge ordering, $\Delta$ – initial half-width of miniband describing interlayer motion of electrons in the disordered crystal.

Substitution of Eq.(3) into formula (2) yields the following final result for the longitudinal Seebeck coefficient:

$$\alpha_{zz} = \pi \alpha_0 \frac{A}{B+C}. \qquad (5)$$

In this formula, $\alpha_0 = k/e$, and the dimensionless values $A, B, C$ depend on the way of simulation of relaxation time dependence on the longitudinal quasi-pulse. In the case of the relaxation time proportional to the longitudinal velocity modulus they are as below:

$$A = \sum_{l=1}^{\infty} (-1)^l f_l^{th} \int_{-(\gamma-b)}^{\sqrt{w^2\delta^2+1}} y^{-2} \left(1 + w^2\delta^2 - y^2\right)\left(y^2 - w^2\delta^2\right) \sin\left[\pi l b^{-1}(\gamma - y)\right] dy, \qquad (6)$$

$$B = 0.5 \int_{-(\gamma-b)}^{\sqrt{w^2\delta^2+1}} y^{-2} \left(1 + w^2\delta^2 - y^2\right)\left(y^2 - w^2\delta^2\right) dy, \qquad (7)$$

$$C = \sum_{l=1}^{\infty} (-1)^l f_l^{\sigma} \int_{-(\gamma-b)}^{\sqrt{w^2\delta^2+1}} y^{-2} \left(1 + w^2\delta^2 - y^2\right)\left(y^2 - w^2\delta^2\right) \cos\left[\pi l b^{-1}(\gamma - y)\right] dy. \qquad (8)$$

In the case of constant relaxation time they are as follows:

$$A = \sum_{l=1}^{\infty}(-1)^l f_l^{th} \int_{-(\gamma-b)}^{\sqrt{w^2\delta^2+1}} |y|^{-1}\sqrt{(1+w^2\delta^2-y^2)(y^2-w^2\delta^2)}\sin[\pi l b^{-1}(\gamma-y)]dy, \qquad (9)$$

$$B = 0.5 \int_{-(\gamma-b)}^{\sqrt{w^2\delta^2+1}} |y|^{-1}\sqrt{(1+w^2\delta^2-y^2)(y^2-w^2\delta^2)}\,dy, \qquad (10)$$

$$C = \sum_{l=1}^{\infty}(-1)^l f_l^{\sigma} \int_{-(\gamma-b)}^{\sqrt{w^2\delta^2+1}} |y|^{-1}\sqrt{(1+w^2\delta^2-y^2)(y^2-w^2\delta^2)}\cos[\pi l b^{-1}(\gamma-y)]dy. \qquad (11)$$

In these formulae, $\gamma = \zeta/\Delta$, $b = \mu^* B/\Delta$, $w = W_0/\Delta$. Besides:

$$f_l^{th} = [\text{sh}(\pi^2 lkT/\mu^* B)]^{-1}[1-(\pi^2 lkT/\mu^* B)\text{cth}(\pi^2 lkT/\mu^* B)], \qquad (12)$$

$$f_l^{\sigma} = \frac{\pi^2 lkT/\mu^* B}{\text{sh}(\pi^2 lkT/\mu^* B)}. \qquad (13)$$

In formulae (12) and (13), sh(…) and cth(…) are the hyperbolic sine and cotangent, respectively.

If $\gamma - b \geq -w\delta$, the low integration limit in formulae (5)-(10) should be replaced by $w\delta$.

Chemical potential $\zeta$ and order parameter $\delta$ in a quantizing magnetic field at low temperatures are found from the following system of equations [11]:

$$\frac{1}{2\pi\zeta_0}\left[\int_{\zeta+R\geq 0}(\zeta+R)dx + \int_{\zeta-R\geq 0}(\zeta-R)dx\right] + \frac{kT}{\zeta_0}\sum_{l=1}^{\infty}\frac{(-1)^l}{\text{sh}(\pi^2 lkT/\mu^* B)}\times$$

$$\times\left[\int_{\zeta+R\geq 0}\sin\left(\pi l\frac{\zeta+R}{\mu^* B}\right)dx + \int_{\zeta-R\geq 0}\sin\left(\pi l\frac{\zeta-R}{\mu^* B}\right)dx\right] = 1, \qquad (14)$$

$$\frac{\delta W_0}{2\pi\zeta_0}\left[\int_{\zeta+R\geq 0}(\zeta+R)R^{-1}dx - \int_{\zeta-R\geq 0}(\zeta-R)R^{-1}dx\right] + \frac{\delta kTW_0}{\zeta_0}\sum_{l=1}^{\infty}\frac{(-1)^l}{\text{sh}(\pi^2 lkT/\mu^* B)}\times$$

$$\times\left[\int_{\zeta+R\geq 0}R^{-1}\sin\left(\pi l\frac{\zeta+R}{\mu^* B}\right)dx - \int_{\zeta-R\geq 0}R^{-1}\sin\left(\pi l\frac{\zeta-R}{\mu^* B}\right)dx\right] = \delta. \qquad (15)$$

In these equations, $R = \sqrt{W_0\delta^2 + \Delta^2 \cos^2 x}$, $\zeta_0 = n_0 a h^2 / 4\pi m^*$ – the Fermi energy of an ideal two-dimensional Fermi-gas with the absolute zero, $n_0$ – volumetric concentration of current carriers. Integration is done by the positive values of $x$.

It is noteworthy that Eq.(15) always has a trivial solution $\delta = 0$ which corresponds to the disordered state. However, at $W_0/\zeta_0 > 1$ the system of equations (14) – (15) has a nontrivial solution. Condition for the existence of a nontrivial solution means that to realize the ordered state, the electron-phonon interaction should not just exceed the Coulomb one, but this excess must be larger than the maximum kinetic energy of interlayer motion of electrons at $T = 0$ in the case of an ideal two-dimensional Fermi gas. Exactly in this case the value of order parameter $\delta$, even close to unity, proves to be energetically favourable, despite the Coulomb repulsion of carriers and increase of kinetic energy of their interlayer motion when passing to the ordered phase. Otherwise, only a disordered state is possible which corresponds to $\delta = 0$.

Comparing this model to the experimental data, we note that when passing to the ordered phase with formation of *interlayer* CDW of transient metal dichalcogenides, such as $2H\text{-}NbSe_2$ [12], and graphite compounds intercalated with alkali metals, such as $C_8Rb$ [13], the anisotropy of effective masses and their electric conductivity increases drastically. At the same time, if charge ordering were interlayer, these parameters would have to drop.

The results of numerical solution of the system of equations (14)-(15) at $\zeta_0/\Delta = 1$, $W_0/\Delta = 1.5$, $kT/\Delta = 0.03$, $0 \leq \mu^* B/\Delta \leq 5$ are depicted in Figs.1,2.

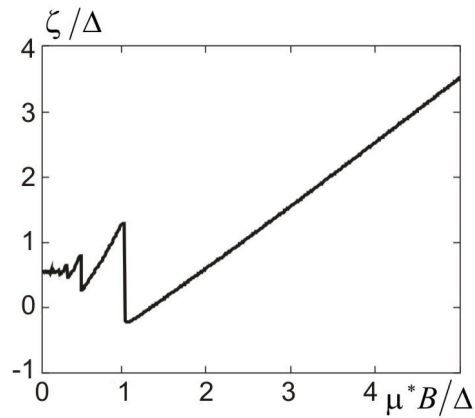

Fig.1. Field dependence of the chemical potential $\zeta$ of the charge-ordered crystal at $\zeta_0/\Delta = 1$, $W_0/\Delta = 1.5$ and $kT/\Delta = 0.03$ in the range of magnetic fields $0 \leq \mu^* B/\Delta \leq 5$.

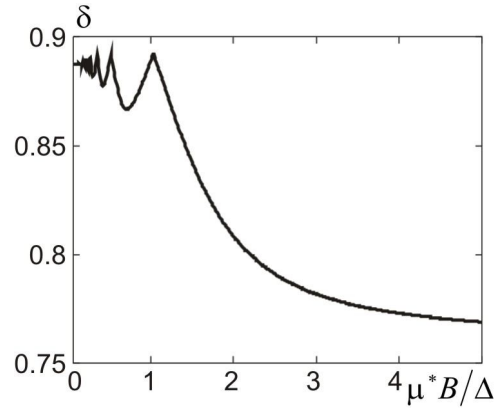

Fig.2. Field dependence of the charge-ordered crystal ordering parameter at $\zeta_0/\Delta = 1$, $W_0/\Delta = 1.5$ and $kT/\Delta = 0.03$ in the range of magnetic fields $0 \leq \mu^* B/\Delta \leq 5$.

From these results it is apparent that the chemical potential and order parameter are the oscillating functions of magnetic field, and that in the strong magnetic fields closer to ultraquantum limit charge ordering is destroyed, though rather slowly. Moreover, even in a weak magnetic field it is not complete, i.e. $\delta < 1$. Charge ordering incompleteness takes place because interlayer motion of electrons whose intensity is characterized by half-width of initial miniband $\Delta$, tends to smooth the nonuniformity of their layer-by-layer distribution. While ordering destruction in a strong magnetic field is due to the fact that kinetic energy of interlayer motion of electrons in the presence of the Landau levels is larger than in the absence of these levels. And kinetic energy increase, from thermodynamic considerations, must resist charge ordering.

With these parameters, the effective mass anisotropy when passing to the ordered state in a weak field increases by a factor of 24, though experiments [12] and [13] give evidence of its much larger growth. This can mean that the values of $\delta$ closer to unity are possible.

From Fig.1 it is seen that in the strong magnetic fields the chemical potential of a system of conduction electrons in the charge-ordered crystal changes in a jump-like fashion. In fact, these jumps are somewhat smeared and smoothed due to the Dingle factor caused by charge carrier scattering on the impurities and crystal lattice defects. However, we will consider, just as it was done in [11], condition $\omega_c \tau \gg 1$ to be fulfilled, where $\omega_c$ − cyclotron frequency, $\tau$ − relaxation time. Exactly under this condition, as it follows from the results of [14], the approaches based on the Kubo formalism and on the kinetic Boltzmann equation, yield identical results. It takes place because under this condition the shift in the energy levels due to scattering can be ignored, and their widening can be directly related to the relaxation time.

The last jump of the chemical potential is matched by a topological transition from an open Fermi surface (FS) to a closed one. Note that in the strict sense in a quantizing magnetic field the FS of crystal is not defined, so it is referred to conventionally, being represented as a combination of hollow coaxial cylinders, the so-called "magnetic tubes" whose axes are parallel to the magnetic field direction and crystal C-axis. In such a case the FS is open, if there is at least one "magnetic tube" occupying (lengthwise) the entire one-dimensional Brillouin zone. Otherwise the FS is considered to be closed.

In the ultraquantum limit the magnetic field dependence of the chemical potential becomes almost linear. In this case it is determined by the following asymptotic formula:

$$\zeta = \mu^* B - \sqrt{W_0^2 \delta^2 + \Delta^2 \cos^2 \frac{\pi \zeta_0}{4\mu^* B}}. \qquad (16)$$

Such form of solution is, in particular, caused by the fact that at $-\sqrt{W_0^2 \delta^2 + \Delta^2} < \zeta - \mu^* B < -W_0 \delta$ the FS of crystal within the first one-dimensional Brillouin zone consists of three "magnetic tubes", one of which has symmetry plane $k_z = 0$, and the other two, being halves of the first tube, start from planes $k_z = \pm \pi/a$ and get narrow inward the Brillouin zone. The relation (15) means that in the strong magnetic fields there is condensation of current carriers toward the bottom of single filled Landau miniband with the number $n = 0$. This causes a topological transition which consists in a gradual conversion of the FS from an open to a closed one. This fact must be necessarily taken into account in the calculation of the longitudinal Seebeck coefficient.

The results of calculation of the longitudinal Seebeck coefficient in the quasi-classical region of magnetic fields, i.e. at $0.04 \leq \mu^* B/\Delta \leq 0.1$ and $\zeta_0/\Delta = 1$, $W_0/\Delta = 1.5$, $kT/\Delta = 0.03$ are depicted in Fig.3.

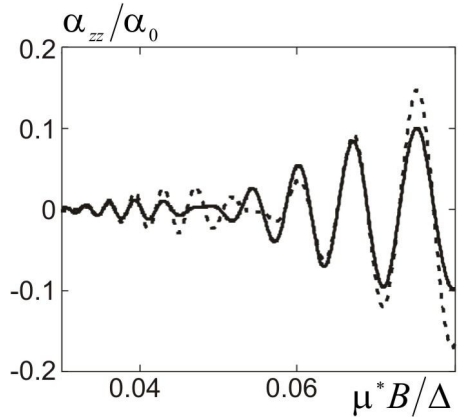

Fig.3. Field dependence of the longitudinal Seebeck coefficient of the charge-ordered crystal at $\zeta_0/\Delta = 1$, $W_0/\Delta = 1.5$ and $kT/\Delta = 0.03$ in the range of magnetic fields $0.04 \leq \mu^* B/\Delta \leq 0.1$. The solid curve is for the relaxation time proportional to the longitudinal velocity, the dashed curve is for constant relaxation time.

In both models, as could be expected, oscillations of the longitudinal Seebeck coefficient have a double periodic structure, the smaller oscillation periods being related to maximum sections of the FS with $k_z = 0$ and $k_z = \pm \pi/a$ planes normal to field. Larger oscillation periods are related to minimum sections of the FS with $k_z = \pm \pi/2a$ planes. The respective oscillation frequencies are determined by the formulae [11]:

$$F_l = \pi l \frac{\zeta + \sqrt{W_0^2 \delta^2 + \Delta^2}}{\mu^*}, \qquad (17)$$

$$F_l' = \pi l \frac{\zeta + W_0 \delta}{\mu^*}. \qquad (18)$$

In the charge-ordered layered crystal these frequencies are close to each other, therefore, oscillations of the longitudinal Seebeck coefficient have the form of high-frequency vibrations with low-frequency beats superimposed on them. As regards the amplitudes and phases of oscillations, they, as can be seen from Fig.3, are essentially dependent on the way of simulation of current carrier relaxation time. In case of the relaxation time proportional to the modulus of the longitudinal velocity, the Seebeck coefficient is on the whole lower than in the case of constant relaxation time. Moreover, in the case of the relaxation time proportional to the modulus of the longitudinal velocity the oscillations lead in phase the oscillations that must be observed in the case of constant relaxation time. Within the main part of oscillations in the investigated range of magnetic fields for both models of the relaxation time the value of the Seebeck coefficient does not exceed $0.157\alpha_0$, i.e. 13.54μV/K. And still, according to our analysis, it is approximately twice that for the same crystal in the absence of charge ordering. Moreover, with selected problem parameters for both models of the relaxation time in the absence of charge ordering the longitudinal Seebeck coefficient is a monotonically increasing function of the magnetic field, and the oscillating function of the magnetic field is only its derivative with respect to the magnetic field. The larger value of the longitudinal Seebeck coefficient as compared to the disordered state is attributable to the fact that without ordering in the quasi-classical region the amplitude of its oscillations is proportional to $(\mu^* B/\Delta)^{3/2}$ with constant relaxation time and to $(\mu^* B/\Delta)^2$ with the relaxation time proportional to the longitudinal velocity. However, in the ordered state this amplitude is proportional to $(\mu^* B/\Delta^*)^{3/2}$ and $(\mu^* B/\Delta^*)^2$, respectively, where $\Delta^* = 0.5(\sqrt{\Delta^2 + W_0^2 \delta^2} - W_0 \delta)-$ half-width of miniband in the presence of charge ordering taking into account that in conformity with (3) there are two such minibands, but only the lower one is filled. Therefore, as long as at $\delta \neq 0$ $\Delta^* < \Delta$, the longitudinal Seebeck coefficient in the ordered state is also higher. From the above it is clear that in the weak magnetic fields at low temperatures the longitudinal Seebeck coefficient is close to zero in conformity with the general principles of thermodynamics.

The field dependence of the longitudinal Seebeck coefficient in a wide range of magnetic fields $0 \leq \mu^* B/\Delta \leq 2$ at $\zeta_0/\Delta = 1$, $W_0/\Delta = 1.5$, $kT/\Delta = 0.03$ is shown in Fig.4.

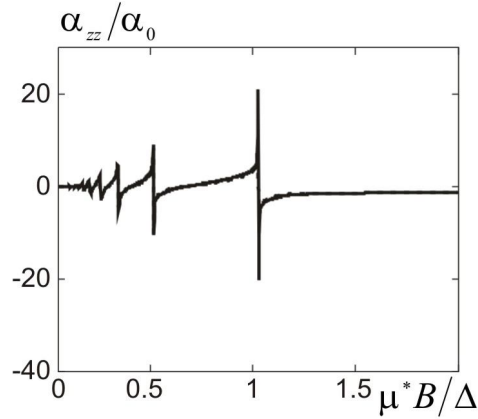

Fig.4. Field dependence of the longitudinal Seebeck coefficient of the charge-ordered crystal at $\zeta_0/\Delta = 1$, $W_0/\Delta = 1.5$ and $kT/\Delta = 0.03$ in the range of magnetic fields $0 \leq \mu^* B/\Delta \leq 2$. The solid curve is for the relaxation time proportional to the longitudinal velocity, the dashed curve is for constant relaxation time.

From this figure it is seen that with magnetic field increase, the value of the Seebeck coefficient increases considerably, and in conformity with the chemical potential jumps its polarity (sign) is reversed. The maximum value of the Seebeck coefficient in this case is achieved at a point of topological transition from an open FS to a closed one which corresponds to the last jump of the chemical potential. This value is about $23.55\alpha_0$, i.e. 2033μV/K. Without charge ordering the maximum value of the Seebeck coefficient in a strong magnetic field with selected problem parameters for both models of the relaxation time does not exceed 118μV/K, and polarity reversal does not take place. Polarity reversal of the Seebeck coefficient in the charge-ordered layered crystal is attributable to openness of its FS. Without charge ordering, with selected problem parameters the FS of a layered crystal is closed. However, drastic increase in the Seebeck coefficient in the charge-ordered state is attributable not only to the chemical potential jump with a topological transition, but also to contraction of conduction miniband due to charge ordering.

After the last polarity reversal, the Seebeck coefficient becomes "purely electron", i.e. negative, and its modulus decays to zero according to a certain asymptotic law. This law can be established as follows. In formulae (4)-(10), with regard to the FS compression along the magnetic field direction, from integration over the dimensionless energy we will pass to integration over the dimensionless quasi-pulse in the region $0 \leq x \leq \pi\zeta_0/4\mu^* B$. In formulae for coefficients $A, B, C$ we substitute the expression for $\zeta$ in conformity with formula (15). Then multipliers $(-1)^l$ will be compensated. The integrands will be expanded as series in small parameters $x$ and $\pi\zeta_0/4\mu^* B$ to the second order of smallness inclusive. Besides, at $\mu^* B/kT \gg 1$ we will represent $f_l^{th}$ as:

$$f_l^{th} = \frac{\left(\pi^2 lkT/\mu^* B\right)^2}{2\operatorname{sh}\left(\pi^2 lkT/\mu^* B\right)}, \qquad (19)$$

and take into account that numerical analysis shows the validity of the following relations (true at low $x$):

$$\sum_{l=1}^{\infty} \frac{xl}{\operatorname{sh}(xl)} = \frac{2.467}{x}, \qquad (20)$$

$$\sum_{l=1}^{\infty} \frac{xl^3}{\operatorname{sh}(xl)} = \frac{12.176}{x^3}. \qquad (21)$$

As a result, we obtain the following asymptotic expression for the modulus of the longitudinal Seebeck coefficient valid in the ultra-quantum limit for both models of the relaxation time:

$$|\alpha_{zz}| = K\alpha_0 \frac{\zeta_0^2 \Delta^2}{kT\left(\mu^* B\right)^2 \sqrt{W_0^2 \delta^2 + \Delta^2}}. \qquad (22)$$

Coefficient of proportionality $K$ is 0.305 for constant relaxation time and 0.254 for the relaxation time proportional to the longitudinal velocity. On the face of it, such asymptotic law is elusive, since the longitudinal Seebeck coefficient does not tend to zero at $T = 0$. However, at real low temperatures this law does not lead to physical incorrect results. Indeed, for instance, at $m^* = 0.01 m_0$, $\Delta = 0.01 \text{эВ}$, $kT/\Delta = 0.03$, $B = 60 \text{Тл}$, $\zeta_0/\Delta = 1$, $W_0/\Delta = 1.5$, $\delta \approx 0.77$ (in conformity with Fig.2, considering that charge ordering is destroyed slowly) for the modulus of the longitudinal Seebeck coefficient we obtain the value 0.474μV/K at constant relaxation time and 0.396μV/K at the relaxation time proportional to the longitudinal velocity. And it is, respectively, a factor of 4285 and 5135 smaller than the longitudinal Seebeck coefficient maximum achieved with a topological transition, if, in conformity with Fig.4 this maximum is considered to be weakly dependent on the choice of charge carrier relaxation time model. The law proposed in this paper has been obtained within the framework of the same approach when the longitudinal electric resistance is proportional to $TB^2$ with constant relaxation time and $TB^3$ with the relaxation time proportional to the longitudinal velocity [11].

This raises the question as to the possibility of obtaining a different asymptotic law of a change in the longitudinal Seebeck coefficient in the ultraquantum limit. If we apply to [14], within the framework of its approach, with a simultaneous fulfillment of conditions $\omega_c \tau \gg 1$, $kT/\mu^* B \ll 1$ and $\Delta/\mu^* B \ll 1$ by virtue of independence of nonequilibrium addition to the distribution function of temperature, the thermal diffusion flux, hence, the longitudinal Seebeck coefficient, must vanish identically. However, one cannot say from which exactly value of magnetic field induction it must happen. Therefore, in the framework of this approach, the approximation used in [14] is sufficient only for the calculation of the longitudinal electric conductivity, which is the case here, while the longitudinal Seebeck coefficient should be calculated in the next approximation by the smallness parameters $kT/\mu^* B$ and $\Delta/\mu^* B$. However, it is not within the scope of the present paper.

**3. Conclusion**

Thus, in this paper it is shown that in the quasi-classical quantizing magnetic fields the oscillations of the longitudinal Seebeck coefficient of the charge-ordered layered crystal have a double-periodic structure caused by the FS openness. In so doing, the amplitude of oscillations in the charge-ordered state is larger than in the disordered state, which is due to contraction of conduction miniband with charge ordering. In the weak quantizing magnetic fields the amplitude of oscillations of the Seebeck coefficient depends on the magnetic field induction by the law $B^{1/2}$ with constant relaxation time and by the law $B^{3/2}$ with the relaxation time proportional to the longitudinal velocity. In both cases the amplitude of oscillations of the longitudinal Seebeck coefficient does not exceed 4.313μV/K.

In the strong quantizing magnetic fields due to the chemical potential jumps there is a drastic rise in the longitudinal Seebeck coefficient and its polarity reversal. The maximum modulus of the longitudinal Seebeck coefficient corresponds to a topological transition from an open FS to a closed one and with selected problem parameters it is equal to 2033μV/K, whereas in the disordered state the maximum modulus of the Seebeck coefficient does not exceed 118μV/K.

As long as in reality the jumps of the chemical potential, hence, of the longitudinal Seebeck coefficient, are smeared due to the Dingle factor caused by charge carrier scattering on impurities and crystal lattice defects, the maximum value of the longitudinal Seebeck coefficient and the abruptness of its polarity reversal in a magnetic field close to the ultraquantum limit can serve as criteria of sample purity and perfection.

In the strong quantizing magnetic fields after the point of a topological transition the modulus of the longitudinal Seebeck coefficient drops to zero by the asymptotic law $|\alpha_{zz}| \propto T^{-1}B^{-2}$ identical for both models of the relaxation time which is due to the FS compression along the direction of the magnetic field with charge carrier condensation toward the bottom of the only filled Landau sub-band with the number $n = 0$. The last-mentioned result cannot be obtained within the framework of conventional quasi-classical approach for which the specific shape of the FS and its extension along the direction of the field are insignificant.

The author hopes that this theoretical study will give an impetus to organization and performance of experimental works aimed at investigating the Seebeck coefficients of layered and quasi-dimensional systems with both strongly open and closed FS in a wide range of quantizing magnetic fields.